\documentstyle[preprint,revtex]{aps}
\newcommand{\beq}{\begin{equation}}
\newcommand{\eeq}{\end{equation}}
\newcommand{\beqar}{\begin{eqnarray}}
\newcommand{\eeqar}{\end{eqnarray}}
\begin{document}
%\draft
\preprint{March 24, 1993 KSUCNR-004-93, CEBAF-TH-05}
\begin{title}
Gauge Invariance and the Electromagnetic Current\\
of Composite Pions
\end{title}
\author{M.R. Frank}
\begin{instit}
Department of Physics, Hampton University
Hampton, Virginia 23668\\
and\\
Continuous Electron Beam Accelerator Facility\\
12000 Jefferson Avenue, Newport News, Va. 23606
\end{instit}
\author{P.C. Tandy}
\begin{instit}
Center for Nuclear Research,
Department of Physics\\
Kent State University,
Kent, Ohio 44242
\end{instit}
\begin{abstract}
The Global Color-symmetry Model of QCD is extended to deal with a
background electromagnetic field and the associated conserved current is
identified for composite $\bar{q}q$ pion modes of the model.
Although the analysis is limited to tree level in the
bilocal fields that bosonize the model, the identified photon-pion vertex
produces the charge form factor associated with ladder Bethe-Salpeter
pion amplitudes.  A Ward-Takahashi identity for this vertex is derived
in terms of the effective inverse propagator for the equivalent local
pion field and the intrinsic ladder Bethe-Salpeter amplitudes.
This identity is then used to illustrate gauge invariance by showing that
identical vertex information is produced from the gauge change of
the free action once proper account is taken of the gauge transformation
properties of the bilocal pion fields.  Comments are made on the
location of the vector dominance mechanism in this treatment.
\end{abstract}

\section {Introduction}

The planned experimental program at the Continuous Electron Beam Accelerator
Facility (CEBAF) will subject the electromagnetic (EM) structure
of hadrons and nuclei to detailed scrutiny.  Effective field theory
descriptions of interacting hadrons usually incorporate the
intrinsic nonlocality through empirical form factors to simulate the
underlying degrees of freedom that are not treated explicitly.  The EM
current for such a model is complicated by the fact that the form
factors can induce additional contributions over and above the
canonical current appropriate to point particles.  Without a description of the
composite hadrons in terms of the constituent charged field degrees of freedom,
the EM current is not uniquely defined.  Gauge invariance and current
conservation as implemented through Ward-Takahashi identities provide
only a partial constraint.  The pion exchange current between nucleons
has received the most attention in this regard.  Prescriptions based
on minimal substitution into momentum variables, including those of
form factors, have been developed\cite{gross87}
to impose such partial constraints.

An ideal perspective on this problem would be provided if the composite
hadron fields and their interactions could be modeled in a manageable
form in terms of the point fields of QCD.  The hadronic EM current
could then, in principle,
be directly related to the bare quark current.
Some progress has been made in recent years towards the
hadronization\cite{cahill89,reinhardt90} of
simplified models of QCD based on a four-fermion interaction. However,
for complex hadronic systems such as nucleons interacting through pion
exchange, these
methods are far from yielding a transparent and realistic quark basis
for the associated EM current.
A much simpler situation where a hadronic EM
current can be generated from the quark level is provided by the meson
sector produced from bosonization of four-quark interaction models of
the Nambu--Jona-Lasinio (NJL) type.\cite{nambu61}  With the usual
contact form of the interaction, the composite $\bar{q}q$ meson modes at
the mean field or tree level are point objects.  To conduct a more realistic
investigation of the relation between the EM currents at the quark and
hadronic level, we employ the finite range generalization known as the
Global Color-symmetry Model (GCM).\cite{praschifka87}  The feature of the
GCM that is of interest here, is that the resulting fields for $\bar{q}q$
mesons have finite extent.  The purpose of this paper is to develop
the conserved
EM current of the extended pions in terms of the quark EM current and to
show how gauge invariance at the level of effective localized pion fields
is realized in the presence of the intrinsic nonlocalities.
It is known that Ward-Takahashi identities are not limited to point
particles.\cite{nishijima}
Here we develop several explicit
forms of a Ward-Takahashi identity for the photon-pion coupling,
and also discuss the role of gauge transformation properties of the
extended pion fields in maintaining gauge invariance of the action.

The NJL model has been developed into a very efficient and accessible
representation of hadron dynamics that is believed to capture the
important elements of low energy QCD phenomena.\cite{vogl91}
In the GCM  model a somewhat more fundamental stance is attempted in that
quark color currents interact via a phenomenological gluon two-point
function which can be modeled to incorporate confinement and
asymptotic freedom aspects of QCD.  The desirable features of hidden
chiral symmetry and dynamical quark mass generation are present, but
only a global color symmetry is implemented.  The required bilocal field
bosonization methods have been developed, and a thorough analysis of the
pseudoscalar octet of Nambu-Goldstone bosons has reproduced
many elements of effective chiral meson actions,
including anomalous terms and current algebra results.\cite{roberts88}
The dynamical self-energy amplitudes and related vertices
introduce  nonlocalities that provide a natural convergence
to the quark loop integrals that govern meson dynamics.
A parameterization of the effective gluon
propagator has been developed that produces  satisfactory results for
meson masses, coupling constants and decays.\cite{mesondata}
We have recently explored a generalization that retains
valence quarks to produce a chiral quark-meson baryon model
at the mean field level.\cite{ftf90}  This model takes advantage of the
fact that chiral symmetry and the axial Ward identity\cite{delbourgo79}
force a single function to describe both the quark scalar self-energy
amplitude and the distributed vertex for coupling of quarks
to the $\sigma$ and $\vec{\pi}$ mesons.
With an absolutely confining property embodied in the quark self-energy
amplitudes, a numerical solution\cite{frank91} in the absence of the pion
gives  sensible
results for many nucleon properties.  This work, with one free parameter,
demonstrates the physical operation and sensible outcome of a nucleon
model where the lack of a vacuum quark mass-shell cooperates with a scalar
bilocal quark condensate induced by nearby valence quarks to produce a
constituent mass-shell.  The
EM coupling to this nucleon model is under investigation,
and will be reported elsewhere.

For the pion charge form factor, the NJL model in mean field theory produces a
behavior that is only qualitatively correct\cite{blin88}
over the time-like and space-like regions spanned by data.
The formulation of the pion charge form factor within the GCM is
of interest in its own right because several dynamical features arise
that are beyond the capability of the contact NJL interaction.
The Bethe-Salpeter pion amplitudes at the mean field level will not be
constants and an extra length scale that contributes to the pion charge
radius will enter this way.  The quark self-energy will be a dynamical
quantity rather than a constant constituent mass, and the associated
photon vertex with the dressed quarks will acquire a quark momentum
dependence.  These aspects are dynamically linked because
in the chiral limit the Dirac scalar component of the quark self-energy
coincides with the pion Bethe-Salpeter amplitude. Finally, the finite
range nature of the GCM can accommodate confinement and spurious
threshold effects from the quark loops should be absent.

In Section II the bosonization of the GCM in the presence of an
external EM field is outlined and involves path integral techniques
that implement the change of field variables from quarks to bilocal
bose fields.  The gauge invariance of the original action is maintained
to produce a gauge invariant EM coupling to the composite pions.  Effective
local fields to describe pion propagation are defined through the
localization procedure of Cahill.\cite{cahill89}  This results in a
factorized representation for the bilocal fields with a well-defined
internal structure factor that becomes the ladder Bethe-Salpeter
amplitude on the mass-shell.  At this stage the effective local pion
fields appear in a nonlocal action in which the EM current and charge
form factor involve the hadronic form factor of the pion.

In Section III we analyze the gauge invariance of the description.
Ward-Takahashi identities are derived for the photon-pion vertex
in terms of the relevant inverse propagators for the
bilocal pion fields and also for the effective local pion fields.
In the latter case, the role
of the intrinsic hadronic amplitudes for the pion are displayed.
The explicit gauge invariance of the action with first-order
photon coupling is analyzed in terms of the transformation
properties of the extended $\bar{q}q$ pion fields.  From this point
of view, the nonlocality of the EM  coupling to extended pions
is shown to generate the longitudinal component of the four-point
vertex associated with pion electro-production on dressed quarks.
The role and location of the vector dominance mechanism in this
approach is discussed briefly in Section IV.  A summary is made in
Section V.

\section{Global Color-symmetry Model}

\subsection{Bosonization in a Background Electromagnetic Field}

We briefly outline the steps necessary to bosonize the partition function
of the GCM in the presence of a background EM field through the
use of bilocal auxiliary fields.  The techniques are a straightforward
generalization of those previously applied\cite{praschifka87,roberts88}
to the GCM in the absence of a background field and we point out the
new elements that arise here.
Although our treatment deals with two quark flavors, for which
the charge operator is
$\hat{Q}=\frac{1}{2}(\tau _{3}+\frac{1}{3})$,
the formalism is easily generalized in that respect.
The action for the GCM with a background EM field can be written in
Euclidean metric as
\begin{eqnarray}
S[\bar{q},q,A_{\nu }]=\int d^{4}xd^{4}y\Biggl\{ \bar{q}(x)\Bigl[
\bigl( \gamma \cdot \partial _{x} +m & - & i\gamma _{\nu }\hat{Q}A_{
\nu}(x)\bigr) \delta(x-y)\Bigr] q(y)\nonumber \\
 & + & \frac{g^{2}}{2}j_{\nu
}^{a}(x)D(x-y)j_{\nu }^{a}(y)\Biggr\} ,\label{2a3}
\end{eqnarray}
where $m$ is a small current quark mass, $A_{\nu}$ describes an external
EM field, and
$j_{\nu }^{a}(x)=\bar{q}(x)\frac{\lambda ^{a}}{2}\gamma _{\nu
}q(x)$ is the quark current.  This action is invariant under the gauge
transformation
\begin{eqnarray}
q(x)\rightarrow q'(x) &=& e^{i\hat{Q}\theta (x)}q(x) , \nonumber \\
\bar{q}(x)\rightarrow \bar{q}'(x) &=& \bar{q}(x)e^{-i\hat{Q}\theta (x)} ,
 \nonumber \\
A_{\nu }(x)\rightarrow A'_{\nu }(x) & = & A_{\nu
}(x)+\partial _{\nu }\theta (x).\label{2a4}
\end{eqnarray}
The phenomenological gluon two-point function $D(x-y)$ is a parameter
function of the GCM and an explicit form that fits low energy meson
dynamics is available.\cite{mesondata}  The details are not important
for the present analysis.
The standard bosonization technique\cite{roberts88,bosonize} can be applied
in the presence of the external EM field to expose bilocal
$\bar{q}q$ fields and their electromagnetic interactions.
The current-current interaction is Fierz-reordered through the identity
\begin{equation}
\left( \frac{\lambda ^{a}}{2}\gamma _{\nu }{\bf 1}_{F}\right) _{ij}
\left( \frac{\lambda ^{a}}{2}\gamma _{\nu }{\bf 1}_{F}\right) _{kl}=
\left( \Lambda ^{\phi }\right) _{il}
\left( \Lambda ^{\phi }\right) _{kj}, \label{2a8}
\end{equation}
where the quantities $\Lambda ^{\phi }$ are given by
\begin{equation}
\Lambda ^{\phi }=\frac{1}{2}\left( {\bf 1}_{D},i\gamma
_{5},\frac{i}{\sqrt{2}}\gamma _{\nu },\frac{i}{\sqrt{2}}\gamma _{\nu
}\gamma _{5}\right) \otimes \left( \frac{1}{\sqrt{2}}{\bf
1}_{F},\frac{1}{\sqrt{2}}\vec{\tau }\right) \otimes \left( \frac{4}{3}{\bf
1}_{C},\frac{i}{\sqrt{3}}\lambda ^{a}\right) ,\label{2a9}
\end{equation}
which we write as $\Lambda^{\phi} = \frac{1}{2} K^{a}F^{b}C^{c}$.
The product of currents in (\ref{2a3}) becomes
\begin{equation}
j_{\nu }^{a}(x)j_{\nu }^{a}(y)= - J^{\phi}(x,y)J^{\phi
}(y,x), \label{2a10}
\end{equation}
where $J^{\phi }(x,y)=\bar{q}(x)\Lambda ^{\phi }q(y)$
and we use a summation convention for repeated indices.
The gauge invariance of the current-current interaction is preserved
by Fierz reordering.  The flavor sum on the right hand side of
(\ref{2a10}) splits into multiplets whose members mix under
gauge transformations.  In general, fields corresponding to a
complete multiplet must be retained for manifest gauge invariance.
An example is the pair of charged pions and we shall eventually
limit our considerations to this case.

Field variables that eventually take up the role of boson modes of the
model are introduced through multiplication of
the partition function, $Z\left[ A_{\nu }\right] =N\int D\bar{q}Dq
e^{-S\left[ \bar{q},q\right] }$, by unity in the form
\begin{equation}
1={\cal N}\int D{\cal B}^{\phi }exp\left\{ - \int d^{4}xd^{4}y\frac{{\cal
B}^{\phi }(x,y){\cal B}^{\phi }(y,x)}{2g^{2}D(x-y)}\right\}.  \label{2a11}
\end{equation}
With a shift of integration variables
${\cal B}^{\phi }(x,y)\rightarrow {\cal B}^{\phi
}(x,y)+g^{2}D(x-y)J^{\phi }(y,x)$,
the current-current term $\sum _{\phi }J^{\phi }(x,y)J^{\phi }(y,x)$
is eliminated from the action in favor of a Yukawa coupling term
$\bar{q}{\cal B}q$ and a quadratic term in ${\cal B}$.
The partition function is then given by
$Z[A_{\nu }]=N\int D\bar{q}DqD{\cal B}e^{-S\left[ \bar{q},q,{\cal B}\right] }$
where the new action is
\beq
S\left[ \bar{q},q,{\cal B},A_{\nu }\right] =\int
d^{4}xd^{4}y\Biggl\{ \bar{q}(x) {\cal G}^{-1}(x,y)  q(y) +
\frac{{\cal B}^{\phi }(x,y){\cal B}^{\phi }(y,x)}{2g^{2}D(x-y)}\Biggr\} ,
\label{2a14}
\eeq
and the inverse quark propagator is
\begin{equation}
{\cal G}^{-1}(x,y)=\left( \gamma \cdot \partial _{x}+m
-i\hat{Q}\gamma _{\nu }A_{\nu }(x)\right) \delta
(x-y)+\Lambda ^{\phi }{\cal B}^{\phi }(x,y). \label{2a18}
\end{equation}
The boson fields ${\cal B}^{\phi }(x,y)$ have the same EM gauge transformation
properties as the bilocal currents $J^{\phi}(y,x)$, and
the bosonized action (\ref{2a14}) is gauge invariant.
In particular, under a gauge transformation it is evident that
\begin{equation}
\left[ \Lambda ^{\alpha}{\cal B}^{\alpha}(x,y)\right] '=e^{i\hat{Q}\theta
(x)}\left[ \Lambda ^{\beta}{\cal B}^{\beta}(x,y)\right]
e^{-i\hat{Q}\theta (y)}. \label{2a15}
\end{equation}
The transformation law can thus be written
\begin{equation}
{\cal B}^{\alpha }(x,y)'= {\Omega}_{\alpha \beta }
(x,y){\cal B}^{\beta }(x,y),\label{2a14a}
\end{equation}
where the transformation matrix ${\Omega}_{\alpha \beta }$ is given by
\begin{equation}
{\Omega}_{\alpha \beta }(x,y) =
tr_{f}\left[ F^{\alpha }e^{i\hat{Q}\theta
(x)}F^{\beta }e^{-i\hat{Q}\theta (y)} \right] ,\label{2a14b}
\end{equation}
if the flavor basis has the orthonormal property
$tr_{f}[F^{\alpha }F^{\beta}]=\delta _{\alpha \beta }$.
These transformation properties reflect the fact that the
bilocal fields are constructed from charged constituents. In the local
limit where ${\cal B}^{\alpha}(x,y)\rightarrow \delta(x-y) b^{\alpha}(x)$,
uncharged meson fields that couple to quarks via flavor matrices ${\bf 1}$ and
$\tau_{3}$ become gauge invariant, while the standard
gauge transformations operate for local charged fields. For bilocal fields
of finite extent, even the uncharged fields will mix and transform under a
gauge
change.  We shall confine our attention to charged pion modes where the
gauge transformation properties signal an
electromagnetic coupling to both the propagation coordinate $R=(x+y)/2$
and the substructure coordinate $r=x-y$.  Later we shall relate
this to the fact that the electromagnetic current for the composite pion
field is the usual point form times the charge form factor.

Since the quark fields now occur only quadratically
in (\ref{2a14}), the bosonization is completed by Grassmann integration
to give
$Z\left[ A_{\nu }\right] =N\int D{\cal B}e^{-S\left[ {\cal
B}^{\phi },A_{\nu }\right] }$
where the gauge invariant boson action is
\begin{equation}
S\left[ {\cal B}^{\phi },A_{\nu }\right] = -TrLn {\cal G}^{-1}\left[
{\cal B}^{\phi },A_{\nu }\right] +  \int d^{4}xd^{4}y\frac{{\cal
B}^{\phi }(x,y){\cal B}^{\phi }(y,x)}{2g^{2}D(x-y)} . \label{2a17}
\end{equation}
The classical field configurations ${\cal B}^{\phi }_{0}$
are defined by $\frac{\delta S}{\delta
{\cal B}^{\phi }_{0}}=0$  and this produces
\begin{equation}
{\cal B}^{\phi }_{0}(x,y)=g^{2}D(x-y)tr\left[ \Lambda ^{\phi }{\cal
G}_{0}(x,y)\right] ,\label{2b2}
\end{equation}
where the associated propagator ${\cal G}_{0}$ depends
self-consistently on ${\cal B}^{\phi }_{0}$. In particular,
\begin{equation}
{\cal G}^{-1}_{0}(x,y)= \left( \gamma \cdot \partial _{x}+m
-i\hat{Q}\gamma_{\nu }A_{\nu }(x)\right) \delta (x-y)
+\Sigma (x,y) ,\label{2b3}
\end{equation}
where
$\Sigma (x,y) = \Lambda ^{\phi }{\cal B}^{\phi
}_{0}(x,y)$ represents the quark self-energy to the present level of
treatment and satisfies
\begin{equation}
\Sigma (x,y)=\frac{4}{3}g^{2}D(x-y)\gamma _{\nu }{\cal G}_{0}(x,y)\gamma
_{\nu }.\label{2b4}
\end{equation}
This is obtained from (\ref{2b2}) by reversing the Fierz reordering and
carrying out the color sum to produce the $\frac{4}{3}$ factor.
Because of the background electromagnetic field, the quark propagator
and self-energy are not translationally invariant.
In the limit \mbox{$A_{\nu}\rightarrow 0$},
$\Sigma$ and ${\cal G}_{0}$ can be chosen to depend only on $x-y$ and
(\ref{2b4}) becomes the ladder Schwinger-Dyson equation with a bare
vertex.  We use the notation
\mbox{$G_{0}(x-y)={\cal G}_{0}(x,y)|_{A_{\nu}=0}$},
and the momentum representation will be denoted by
\mbox{$G_{0}^{-1}(q)=i\gamma \cdot q A(q^2)+m+B(q^2)$}.
Only the first-order dependence of
${\cal G}_{0}^{-1}$ upon $A_{\nu}$ is of interest and this will generate
the EM vertex with dressed quarks.

The action may be expanded about the classical minimizing configuration
to obtain
\begin{equation}
S\left[{\cal B}^{\phi },A_{\nu }\right]
= S\left[ {\cal B}^{\phi }_{0},A_{\nu }\right] + \hat{S}\left[
\hat{{\cal B}}^{\phi },A_{\nu }\right] , \label{2b17}
\end{equation}
where $\hat{{\cal B}}^{\phi }={\cal B}^{\phi }
-{\cal B}^{\phi }_{0}$ are the new field variables for the propagating
modes and the corresponding action is
\begin{equation}
\hat{S}\left[ \hat{{\cal B}}^{\phi },A_{\nu }\right] = Tr \sum
_{n=2}^{\infty }\frac{(-1)^{n}}{n}\left( {\cal G}_{0}\Lambda^{\phi}
\hat{{\cal B}}^{\phi}\right) ^{n}
+ \int d^{4}xd^{4}y\frac{\hat{{\cal B}}^{\phi }(x,y)
\hat{{\cal B}}^{\phi }(y,x)}{2g^{2}D(x-y)}.\label{2b18}
\end{equation}
Although only color singlet components are present in the classical
configurations ${\cal B}_{0}^{\phi}$, the fluctuation fields of the
GCM have color singlet and octet components. There are indications
that a proper role for the color octet sector is realized through
a transformation that introduces diquark fields for a description
of baryons.\cite{cahill89}
The electromagnetic current is to be identified from
\begin{equation}
J_{\nu}(x) =
- \left. \frac{\delta S\left[{\cal B}^{\phi },A_{\nu }\right] }{\delta
A_{\nu }(x)}\right| _{A_{\nu }=0} = -\left.
\frac{\delta \hat{S}\left[\hat{{\cal B}}^{\phi},A_{\nu}\right]}{\delta
A_{\nu}(x)}\right| _{A_{\nu}=0}
.\label{2b1}
\end{equation}
The second equality in (\ref{2b1}) expresses the fact that the saddle
point action does not contribute to the current.  The functional
dependence of $S\left[{\cal B}_{0},A_{\nu}\right]$ upon $A_{\nu}$
that enters implicitly through ${\cal B}_{0}\left[A_{\nu}\right]$
makes no contribution since $\frac{\delta S}{\delta {\cal B}_{0}} = 0$.
The explicit $A_{\nu}$ dependence that enters through the bare
coupling term of ${\cal G}_{0}^{-1}$ produces a current
$-tr \left(G_{0}(x,x)\hat{Q}\gamma_{\nu}\right)$ which vanishes due
to symmetric integration.  The saddle point action can contribute
to higher order in $A_{\nu}$ beginning with a vacuum polarization
insertion for the photon propagator.  We here ignore such phenomena
since we seek only the current to which a background electromagnetic
field couples.  The saddle point action at $A_{\nu}=0$ is a constant
and is absorbed into the normalization of the partition function
which now is
\begin{equation}
Z\left[ A_{\nu }\right] = N\int D\hat{{\cal B}} e^{-\hat{S}\left[
\hat{{\cal B}}^{\phi }\right] +\int d^{4}x J_{\nu }(x)A_{\nu }(x) +
\ldots },\label{2b21}
\end{equation}
where $J_{\nu}$ is the current for the bosonized action $\hat{S}
[\hat{{\cal B}}]$.

Our considerations are restricted to pion modes and we adjust the
normalization of the fields so that
\begin{eqnarray}
\Lambda^{\phi}\hat{{\cal B}}^{\phi}(x,y) &=& i\gamma_{5}\vec{\tau}\cdot
\vec{\pi}(x,y) \\
&=& i\gamma_{5} f^{\dagger}_{\alpha} \pi_{\alpha}(x,y)
, \label{pifld}
\end{eqnarray}
where $\alpha$ is a summation index that takes the values $\alpha = (0,+,-)$
corresponding to a spherical isospin basis.  This basis is convenient because
it diagonalizes the quark charge operator $\hat{Q}$.  We choose $f_{0}=
\tau_{3}$, $f_{+}=(\tau_{1}+i\tau_{2})/\sqrt{2}$ so that $tr_{f}(f_{\alpha}
^{\dagger}f_{\beta})=2\delta_{\alpha \beta}$.  The corresponding field
components $\pi_{\alpha}$ are defined similarly.   If the flavor
basis is hermitian, the
bilocal fields are hermitian, that is, $\left[{\cal B}^{\phi}(x,y)\right]
^{*} = {\cal B}^{\phi}(y,x)$.  In terms of the basis we have chosen, the
corresponding property is $\left[\pi_{+}(r;R)\right]^{*} = \pi_{-}(-r;R)$
and $\left[\pi_{0}(r;R)\right]^{*} = \pi_{0}(-r;R)$ where $r=x-y$ and
$R=(x+y)/2$.  We use these coordinates to describe internal dynamics and
propagation respectively. In the momentum representation that we shall
use, $(q,P)$ are
conjugate to $(r,R)$, and we have $\left[\pi_{+}(q;P)\right]^{*}=
\pi_{-}(q;-P)$ and $\left[\pi_{0}(q;P)\right]^{*} = \pi_{-}(q;-P)$.

The free action is the quadratic term of (\ref{2b18}) in the absence
of the EM field.  With the above definitions it can be written as
\beq
\hat{S}_{2}\left[\pi \right] = \frac{1}{2}\sum_{\alpha}\int
d^{4}(P,q',q) \pi_{\alpha}^{*}(q';P) \Delta^{-1}(q',q;P) \pi_{\alpha}
(q;P) , \label{S2}
\eeq
where the composite pion inverse propagator is
\beq
\Delta^{-1}(q',q;P) = \delta(q'-q)tr\left[G_{0}(q_{-})i\gamma_{5}
f_{\alpha}G_{0}(q_{+})i\gamma_{5}f_{\alpha}^{\dagger}\right] +
\frac{9}{2}\int \frac{d^{4}r}{(2\pi)^{4}}\frac{e^{-i(q'-q)\cdot r}}
{g^{2}D(r)},\label{delinv}
\eeq
with $tr$ denoting a trace over spin, flavor and color.
Here $q_{+}=q+P/2$ and $q_{-} = q-P/2$ where $P$ is the total momentum
of the meson field and $q$ is the internal momentum associated with the
$\bar{q}q$ substructure.  The first term of the inverse propagator
(\ref{delinv}) is a quark loop and the second term is a bare mass.

\subsection{Localization}

The bilocal fields represent a combination of internal dynamics and
propagation dynamics that is difficult to deal with at the same time.
A useful approach is to use the free action to
define free meson modes in terms of which the bilocal fields may be expanded.
This amounts to a determination of internal form factors for
the composite mesons leaving an associated local field degree of freedom that
describes propagation only.
We are following here the localization procedure developed
by Cahill\cite{cahill89}.  The natural description of free meson modes
is through eigenfunctions defined by
\beq
\int d^{4}q\Delta^{-1}(q',q;P)\Gamma_{n}(q;P)=
\lambda _{n}(P^{2})\Gamma_{n}(q';P),\label{2b21a}
\eeq
where $\lambda_{n}(P^{2})$ is the eigenvalue for the $n{th}$ mode.
The operator is hermitian and the eigenvalues are real.
When the total momentum is such that the eigenvalue vanishes,
(\ref{2b21a}) becomes a free equation of motion, and the
corresponding $\Gamma_{n}$ are the internal form factors for the
freely propagating mesons.  The condition $\lambda_{n}(P^{2}=
-M_{n}^{2})=0$ thus defines the mass-shell and also suggests that
$\lambda_{n}(P^{2})$ provides the appropriate basis for a
composite generalization
of the elementary inverse propagator $P^{2}+M_{n}^{2}$.  This is
the sense in which one has achieved a localization of the problem.
The eigenfunctions $\Gamma_{n}$ form a complete orthogonal set
in terms of which the bilocal fields may be expanded as
\beq
\pi_{\alpha}(q;P)= \sum_{n} \Gamma_{n}(q;P)\pi_{\alpha,n}(P)
,\label{piloc}
\eeq
where $\pi_{\alpha,n}(P)$ are the local effective field variables defined
by projection.

We shall truncate
the expansion to a single term corresponding to the lowest mass $m_{\pi}$.
At $P^{2}=-m_{\pi}^{2}$ the on-mass-shell equation
$\int d^{4}q \Delta^{-1}(q',q;P) \Gamma(q;P)=0$
can be shown, with the help of (\ref{delinv}), to be the ladder Bethe-Salpeter
equation for the pionic $\bar{q}q$ bound state.\cite{praschifka87}
In the chiral ($m_{\pi}=0$) limit this coincides with
the ladder Schwinger-Dyson equation for the Dirac scalar component of the
quark self-energy, that is, $\Gamma(q;0) \propto B(q^2)$.  We choose to
normalize the pion-$\bar{q}q$ amplitudes so that they are dimensionless
and have the chiral limit $\Gamma(q;0) \rightarrow B(q^2)/f_{\pi}$.  The
scale factor $f_{\pi}$ is determined as described below.
For just the ground state modes, the free action can be written
\beq
\hat{S}_{2}[\pi] = \frac{1}{2} \sum_{\alpha}\int d^{4}P
\pi_{\alpha}^{*}(P)\Delta^{-1}(P^{2})\pi_{\alpha}(P) , \label{Sfr}
\eeq
where $\Delta^{-1}$ is the effective inverse propagator defined by
\beqar
\Delta^{-1}(P^{2}) &=& \lambda(P^{2})\int d^{4}q
\Gamma^{*}(q;P)\Gamma(q;P) \nonumber \\
&=& \int d^{4}(q',q) \Gamma^{*}(q';P)\Delta^{-1}(q',q;P)\Gamma(q;P)
. \label{DEL}
\eeqar

Since the eigenvalue becomes zero on the mass-shell, we have
$\Delta^{-1}(P^{2}) = (P^{2}+m_{\pi}^{2})Z(P^{2})$.
The scale factor $f_{\pi}^{-1}$ present in each vertex $\Gamma(q;P)$
is determined so that  $Z$ is unity on the mass-shell and thus
the fields $\pi_{\alpha}(P)$ are physically normalized.
To first order in the current quark mass, the PCAC result
$f_{\pi}^{2}m_{\pi}^{2}=-m<\bar{q}q>$ is reproduced by such an
analysis.\cite{mesondata,roberts88}

The localization procedure does not ignore meson substructure but rather
produces a dynamically equivalent formulation
in terms of local field variables.  The hadronic form factors
$\Gamma(q;P)$ then enter into the coupling of the effective local
pion fields to other fields.  We use this simple but well-defined structure
to explore the role of hadronic form factors in a
gauge invariant EM coupling.
It should be emphasized that the meson fields at this level are bare
or tree-level fields in the sense that quantum dressing effects from
the cubic and higher order couplings among bilocal boson fields
have not been applied.  Nevertheless there is significant dynamical
content in these bare fields as evidenced by the ladder Bethe-Salpeter
structure of the internal form factors.  Such dynamics can provide
a realistic modeling of the pion charge form factor.\cite{gross}
However the dynamical relations between the photon-quark vertex,
the quark self-energy dressing and the Bethe-Salpeter amplitudes
that implement gauge invariance are not easily identified
without a unified development.

\subsection{The Pion Electromagnetic Vertex}

The pion electromagnetic current is
\beq
J_{\nu}(Q) =
- \left. \frac{\delta \hat{S}_{2}\left[\pi,A_{\nu }\right] }{\delta
A_{\nu }(Q)}\right| _{A_{\nu }=0} , \label{JQ}
\eeq
where the quadratic pion term of the action (\ref{2b18}) is to be used.
The electromagnetic field  occurs only in the quark propagator
${\cal G}_{0}$, and
for the quark-photon vertex that arises, we use the definitions
\beqar
\Gamma_{\nu}(p,k;Q) &=& \left. (2\pi)^{2}\frac{\delta {\cal G}_{0}^{-1}(p,k)}
{\delta A_{\nu}(Q)} \right|_{A_{\nu}=0} \nonumber \\
&=& \delta(p-k-Q)\hat{Q} \Gamma_{\nu}\left(\frac{p+k}{2};Q\right)
.\label{qver}
\eeqar
It is then straightforward to obtain the current in the form
\beq
J_{\nu}(Q)=\frac{1}{(2\pi)^{2}} \sum_{\alpha}Q_{-\alpha}
\int d^{4}P
\pi_{\alpha}^{*}(P+\frac{Q}{2})\Lambda_{\nu}(P;Q)\pi_{\alpha}(P-
\frac{Q}{2}), \label{J1}
\eeq
where the integration over the
internal $\bar{q}q$ degree of freedom has been carried out to form the
photon-pion vertex as
\beq
\Lambda_{\nu}(P;Q)= \int d^{4}q \Gamma^{*}
(q+\frac{Q}{4};P+\frac{Q}{2})
\Lambda_{\nu}(P,q;Q)\Gamma(q-\frac{Q}{4};P-\frac{Q}{2}). \label{piver}
\eeq
The quantity $\Lambda_{\nu}(P,q;Q)$ is the photon vertex for bilocal pion
fields and is given by the quark loop expression
\beq
\Lambda_{\nu}(P,q;Q)=tr\left\{G_{0}(q_{-})i\gamma_{5}G_{0}
(q_{+}+\frac{Q}{2})
\Gamma_{\nu}(q_{+};Q)G_{0}(q_{+}-\frac{Q}{2})i\gamma_{5}\right\}
,\label{J2}
\eeq
where $q_{+}=q+\frac{P}{2}$ and $q_{-}=q-\frac{P}{2}$.  The photon-pion
vertex arrived at here at tree level is the impulse approximation with
ladder Bethe-Salpeter amplitudes and is illustrated in Fig. 1.
The charge factors appearing in (\ref{J1}) are derived from the charge
operator according to
\beq
Q_{-\alpha}\delta_{\alpha \beta}=\frac{1}{2}tr_{f}\left[f_{\alpha}
\hat{Q}f_{\beta}^{\dagger}\right] .\label{q}
\eeq
Only the terms involving $Q_{+}$ and $Q_{-}$ will survive
and these are the charges of the $u$ and $d$ quarks respectively.
There are several symmetries obtainable from (\ref{J2}) that are
useful.  The property $\Lambda_{\nu}(-P,-q;Q)=-\Lambda_{\nu}(P,q;Q)$
follows
from charge conjugation, and when this is combined with a $\gamma_{5}$
transformation, the property $\Lambda_{\nu}(P,q;-Q)=
\Lambda_{\nu}(p,q;Q)$ results.
The immediate consequence for the photon-pion vertex is $\Lambda_{\nu}
(P;Q)=-\Lambda_{\nu}(-P;Q)=\Lambda_{\nu}(P;-Q)$.  Thus the $\alpha=0$
term of the pion current (\ref{J1}) vanishes because the integrand
is odd in $P$.  This is the familiar result that the electromagnetic field
does not couple  in first order to a self-conjugate meson.
We obtain this here for composites because the eigenfunction expansion
for the internal structure has been truncated at
the ground state level. Were internally excited meson modes to be
included, transition currents connecting different internal modes would arise.

For the remaining charged fields in (\ref{J1}) we use the standard
notation $\pi(P)=\pi_{+}(P)=\pi_{-}^{*}(-P)$.
With use of the symmetry property
that $\Lambda_{\nu}(P;Q)$ is odd under reversal of $P$, the
quark and antiquark terms may be combined to express
the current in the standard form
\beq
J_{\nu}(Q)=\frac{q_{\pi^{-}}}{(2\pi)^{2}}\int d^{4}P \pi^{*}(P+\frac{Q}{2})
\Lambda_{\nu}(P;Q)\pi(P-\frac{Q}{2}), \label{J3}
\eeq
where $q_{\pi^{-}} = Q_{-}-Q_{+}$ is the $\pi^{-}$ charge.
The field $\pi(P)$ describes a $\pi^{-}$ particle incoming
with momentum $P$ or a $\pi^{+}$ particle outgoing with momentum $-P$.
The previously mentioned momentum symmetries allow the vertex to
to be written $\Lambda_{\nu}(P;Q)
=2 P_{\nu} F_{\pi}+2Q_{\nu} P \cdot Q H_{\pi}$
where both $F_{\pi}$ and $H_{\pi}$ are invariant functions of the
three variables $(P^{2},Q^{2},(P \cdot Q)^{2})$.
For on-mass-shell initial and final pions, we have $(P+\frac{Q}{2})^{2}
=(P-\frac{Q}{2})^{2} =-m_{\pi}^{2}$ or equivalently $P^{2}+\frac{Q^{2}}{4}
=-m_{\pi}^{2}$ and $P \cdot Q=0$.  In that case the surviving vertex
$2P_{\nu}F_{\pi}(Q^{2})$ is purely transverse and $F_{\pi}(Q^{2})$
is the charge form factor compatible with ladder Bethe-Salpeter pions.
The on-mass-shell current is always conserved so long as the vertex
is calculated from (\ref{piver}) and (\ref{J2}) in such a way as to
preserve the important symmetry that $\Lambda_{\nu}(P;Q)$ is odd in
$P$ and even in $Q$.  In particular this will follow, if the relation
(\ref{qver}) between the photon-quark vertex $\Gamma_{\nu}$ and the
quark propagator is maintained.

The point meson limit is obtained by use of the
$Q=0$ value of the vertex and it can be verified that on the mass shell
$\Lambda_{\nu}(P;Q=0)=2P_{\nu}$ and so $F_{\pi}(0)=1$.
Details are given in the Appendix. Thus on the mass-shell,
the composite current contains the point current as a factor, and can
be written
\beq
J_{\nu}(Q)=j_{\nu}(Q) F_{\pi}(Q^{2}), \label{Joms}
\eeq
where the point current in position space is the standard result
\beq
j_{\nu}(R)= -iq_{\pi^{-}}\left[\pi^{*}(R)\partial_{\nu}\pi(R)
-\pi(R)\partial_{\nu}\pi^{*}(R)\right]. \label{jpt}
\eeq

\section{Gauge Invariance}

The action that has been developed
to first-order in the field $A_{\nu}$ and to second-order in the localized
pion field $\pi(P)$ is
\beqar
\hat{S}_{2}[\pi,A] &=& \int d^{4}P \pi^{*}(P)
\Delta^{-1}(P^{2})\pi(P) \nonumber \\
&-& \frac{q_{\pi^{-}}}{(2\pi)^{2}}\int d^{4}(P,Q) \pi^{*}(P+\frac{Q}{2})
\Lambda_{\nu}(P;Q)A_{\nu}(Q)\pi(P-\frac{Q}{2}). \label{act}
\eeqar
The explicit gauge invariance can be exhibited through the development
of a Ward-Takahashi identity. This should relate the longitudinal vertex
to the free inverse propagator so that the gauge change induced in the
free action cancels the first order gauge change of the current term.

\subsection{Ward-Takahashi Identities}

At the photon-quark level, the vertex defined by (\ref{qver}) satisfies
the integral equation
\beq
\Gamma_{\nu }(q;Q) =-i\gamma _{\nu
}-\frac{4}{3}g^{2}
\int \frac{d^{4}k}{(2\pi )^{4}}D(q-k)\gamma _{\mu
}G_{0}\left( k+\frac{Q}{2}\right) \Gamma _{\nu }(k;Q) G_{0}
(k-\frac{Q}{2}) \gamma _{\mu }. \label{3a1}
\eeq
This is easily obtained from the definitions (\ref{2b3}) and (\ref{2b4})
for the vacuum quark propagator and self-energy respectively.  The
Ward-Takahashi identity satisfied by this vertex is
\beq
Q_{\nu }\Gamma _{\nu }(q;Q)=G_{0}^{-1}\left( q-\frac{Q}{2}\right) -
G_{0}^{-1}\left( q+\frac{Q}{2}\right)  . \label{3a2}
\eeq
Since the photon vertex with bilocal pion fields is given by (\ref{J2})
in terms of $\Gamma_{\nu}$, a Ward-Takahashi identity for photon-pion
coupling is readily obtained in the form
\beq
Q_{\nu}\Lambda_{\nu}(P,q;Q)\delta(q'-q) = \Delta^{-1}\left(q'+\frac{Q}{4},
q+\frac{Q}{4};P_{+}\right) - \Delta^{-1}\left(q'-\frac{Q}{4},q-\frac{Q}{4};
P_{-}\right) , \label{3a3}
\eeq
where $P_{+}=P+\frac{Q}{2}$ and $P_{-}=P-\frac{Q}{2}$ are pion total momenta
and $\Delta^{-1}(q',q;P)$ is the inverse propagator for bilocal pion
fields given in (\ref{delinv}).
For localized pion fields, the relevant
vertex is the expectation value of $\Lambda_{\nu}(P,q;Q)$ with respect to
the internal pion form factors according to (\ref{piver}).  The resulting
Ward-Takahashi identity is
\beqar
Q_{\nu}\Lambda_{\nu}(P;Q) = && \int d^{4}(q',q)
\Gamma^{*}(q';P_{+})\nonumber \\
&& \times \left[\Delta^{-1}\left(q',q+\frac{Q}{2};P_{+}\right) -
\Delta^{-1}\left(q'-\frac{Q}{2},q;P_{-}\right)\right]\Gamma(q;P_{-})
. \label{3a4}
\eeqar
This result can be brought closer to a form involving the difference of
effective inverse propagators for localized fields by recognizing that
the form factors in (\ref{3a4}) are eigenfunctions of the propagators
therein. With use of (\ref{2b21a}), the identity (\ref{3a4}) becomes
\beq
Q_{\nu}\Lambda_{\nu}(P;Q)=\left\{ \lambda(P_{+}^{2})-
\lambda(P_{-}^{2})\right\} \int d^{4}q \Gamma^{*}
(q+\frac{Q}{4};P_{+}) \Gamma(q-\frac{Q}{4};P_{-})  . \label{3a5}
\eeq
The right hand side of this Ward-Takahashi identity
should be contrasted with the purely local result
$\Delta^{-1}(P_{+}^{2})-\Delta^{-1}(P_{-}^{2})$ appropriate to elementary
fields.  In the present case, the effective inverse propagator
$\Delta^{-1}$ for the localized factor $\pi(P)$ of the bilocal pion
field contains field substructure information, namely the pion-$\bar{q}q$
vertices as
shown in (\ref{DEL}).  Half of the momentum transfer from the photon
is taken up by the internal structure of the pion and the
final factor in (\ref{3a5}) reflects this.
If the Bethe-Salpeter amplitudes $\Gamma(q;P)$
are everywhere replaced by their on-mass-shell limits, the result
can be written as
\beq
Q_{\nu}\Lambda_{\nu}(P;Q)=\left\{ \Delta^{-1}(P_{+}^{2})-
\Delta^{-1}(P_{-}^{2})\right\} \int d^{4}q \hat{\Gamma}^{*}
(q+\frac{Q}{4}) \hat{\Gamma}(q-\frac{Q}{4})  . \label{3a6}
\eeq
This corresponds to the form expected for a point pion supplemented by
a substructure form factor.  In (\ref{3a6}),
$\hat{\Gamma}$ is normalized so that $\int d^{4}q \hat{\Gamma}^{*}
(q)\hat{\Gamma}(q)=1$, and the final factor in (\ref{3a6}) is a
dimensionless probability amplitude for momentum transfer $\frac{Q}{2}$
to be taken up by the pion internal
wave function.   Further use of the
on-mass-shell limit, shows that the difference of inverse propagators
in (\ref{3a6}) becomes $2P \cdot Q$ and the standard point limit
of the vertex at $Q=0$ is apparent.  For $Q \neq 0$, only the
longitudinal vertex is accessible this way, and the pion charge form
factor cannot be isolated by considerations of gauge invariance alone.
It is clear that the on-mass-shell longitudinal vertex contains only
the dynamics of the Bethe-Salpeter amplitudes and
does not contain the effects of vector-meson dominance or possible
threshold effects associated with the quark loop.

\subsection{Gauge Transformation of the Free Action}

The explicit gauge invariance of the action (\ref{act}) can now be
demonstrated.  Under the infinitesimal change $\delta A_{\nu}(x)=\partial
_{\nu}\theta(x)$, or equivalently, $\delta A_{\nu}(Q)=iQ_{\nu}\theta(Q)$,
the change in the $J \cdot A$ term of the action can be related to free
propagation quantities by using the Ward-Takahashi
identity (\ref{3a5}) for $Q_{\nu}\Lambda_{\nu}(P;Q)$.  If, in turn,
the gauge change in the free action is computed from $\delta \pi(R)=
iq_{\pi^{-}}\theta(R)\pi(R)$, or equivalently $\delta\pi(P)=iq_{\pi^{-}}
\int d^{4}K \theta(P-K)\pi(K)$, there will be no cancellation.  The resolution
of this problem is that the effective inverse propagator $\Delta^{-1}$ has
field content that also transforms under a gauge transformation.  The bilocal
fields have been factorized as $\pi(q;P)=\Gamma(q;P)\pi(P)$ with the internal
amplitudes forming part of the inverse propagator for the field $\pi(P)$.
The gauge transformation properties of the bilocal pion field are not
simply those of $\pi(P)$.

The gauge change induced in the internal factor
of the field can be deduced from the relations (\ref{2a14a}) and (\ref{2a14b}).
For the charged pions, the complex field $\pi(x,y)$ does not mix with
any other field under the gauge transformation, and we have
\beq
\pi(x,y)' = \Omega(x,y) \pi(x,y) , \label{3b1}
\eeq
where
\beqar
\Omega(x,y) &=& e^{iQ_{-}\theta(x)} e^{-iQ_{+}\theta(y)} \nonumber \\
&=& e^{iq_{\pi^{-}}\theta(R)} \omega(r;R) . \label{3b2}
\eeqar
Here $Q_{+}$ and $Q_{-}$ are quark charges, and the second equality
has identified the phase factor containing the pion charge
$q_{\pi^{-}}=Q_{-}-Q_{+}$ that is appropriate for a local field.  The
remaining factor $\omega(r;R)$ implements the extra gauge change that
occurs because of the nonlocality and is
\beq
\omega(r;R) = e^{iQ_{-}[\theta(x)-\theta(R)]} e^{-iQ_{+}[\theta(y)-
\theta(R)]}. \label{3b3}
\eeq
The local limit is $\omega(0;R)=1$.  The first order gauge change in the
bilocal pion field is
\beqar
\delta \pi(r;R) &=& \Bigl\{ iQ_{-}\theta(x) -iQ_{+}\theta(y) \Bigr\}
\pi(r;R) \nonumber \\
&=& \Bigl\{ iq_{\pi^{-}}\theta(R) +iQ_{-}\left[\theta(x)-
\theta(R)\right] -iQ_{+}\left[\theta(y)-\theta(R)\right] \Bigr\} \pi(r;R)
. \label{3b4}
\eeqar
In a momentum representation, and with the original pion field factorized
into the internal $\bar{q}q$ amplitude and the localized field, the first
relation from (\ref{3b4}) becomes
\beq
\delta \pi(q;P) = -\frac{i}{(2\pi)^{2}} \int d^{4}Q \theta(Q) \left\{
Q_{+}\Gamma(q+\frac{Q}{2};P-Q)-Q_{-}\Gamma(q-\frac{Q}{2};P-Q) \right\}
\pi(P-Q) . \label{3b5}
\eeq
An alternative expression in which the purely local result is isolated
as a separate term can be obtained from (\ref{3b5}) through Taylor
expansions in the first argument of the amplitudes $\Gamma$ about the
value $q$.  The result is the momentum representation of the second
of the relations (\ref{3b4}), and is
\beq
\delta \pi(q;P) = \frac{1}{(2\pi)^{2}} \int d^{4}Q \left\{ iq_{\pi^{-}}
\theta(Q)\Gamma(q;P-Q)
+ \delta A_{\nu}(Q) W_{\nu}(q;P,Q) \right\} \pi(P-Q)
. \label{3b6}
\eeq
Here the new quantity introduced in the second term is
\beq
W_{\nu}(q;P,Q)=-\bigl[Q_{+}d_{\nu}(q,Q)+Q_{-}d_{\nu}(q,-Q)
\bigr] \Gamma(q;P-Q) , \label{3b7}
\eeq
where
\beq
d_{\nu}(q,Q)= \left\{\frac{1}{2}\int^{1}_{0} d \eta
e^{\eta\frac{Q}{2} \cdot \frac{\partial}{\partial q} } \right\}
\frac{\partial}{\partial q_{\nu} }  . \label{3b8}
\eeq
All powers of derivatives with respect to the relative $\bar{q}-q$
momentum $q$ have been collected into $W_{\nu}$ together with
the accompanying powers of the photon momentum $Q$ except for one
factor of $Q_{\nu}$ which is
used to identify $\delta A_{\nu}(Q)=iQ_{\nu} \theta(Q)$.
In position space, these developments are equivalent to Taylor expansions of
$\theta(x)$ and $\theta(y)$ about $\theta(R)$ in (\ref{3b4}).

The amplitude $W_{\nu}$ enters only because the
pion has size in that the form factor $\Gamma(q;P)$ supports derivatives
with respect to the internal momentum $q$.  It is a necessary consequence
of gauge transformations on the quark-antiquark field content
and modifies the
internal $\bar{q}q$ vertex of the pion to accommodate
coupling to the (longitudinal) $A_{\nu}$ field generated by a gauge change.
The corresponding physical process is electro-absorption of a pion on a
quark as illustrated in Fig. 2.
Only $Q_{\nu} W_{\nu}(q;P,Q)$ contributes to the field change
(\ref{3b6}) and thus only the longitudinal component of the
four-point electro-absorption vertex enters the present discussion
of gauge invariance.  That information is more conveniently represented
in (\ref{3b5}).

Under an infinitesimal gauge transformation, the free action changes by
\beq
\delta \hat{S}_{2}[\pi] = \int d^4 P \lambda(P^2) \left\{f(P)+f^{*}(P)
\right\} , \label{3b9}
\eeq
where
\beq
f(P) = \int d^4 q \pi^{*}(q;P) \delta \pi(q;P) . \label{3b10}
\eeq
Use of (\ref{3b5}) for $\delta \pi$ produces
\beq
f(P)=\frac{i q_{\pi^{-}}}{(2\pi)^{2}}\int d^{4}Q \theta(Q)\pi^{*}(P)
\pi(P-Q) \int d^4 q \Gamma^{*}(q+\frac{Q}{4};P)
\Gamma(q-\frac{Q}{4};P-Q) . \label{3b11}
\eeq
In obtaining this form, the two terms in (\ref{3b5}) weighted by quark charges
have been combined into a single term weighted by $q_{\pi^{-}}=Q_{-}-Q_{+}$
through
a shift of the integration variable $q$ and by use of the property that
$\Gamma(q;P)$ is even in $q$.  In combination with the $f^{*}(P)$ contribution,
the change in the free action becomes
\beq
\delta \hat{S}_{2}[\pi] = \frac{i q_{\pi^{-}}}{(2\pi)^{2}}\int d^{4}(P,Q)
\pi^{*}(P_{+}) \theta(Q) V(P;Q) \pi(P_{-}) , \label{3b12}
\eeq
where $P_{+}=P+\frac{Q}{2}$ and $P_{-}=P-\frac{Q}{2}$, and
\beq
V(P;Q)=\left\{ \lambda(P_{+}^{2})-
\lambda(P_{-}^{2})\right\} \int d^{4}q \Gamma^{*}
(q+\frac{Q}{4};P_{+}) \Gamma(q-\frac{Q}{4};P_{-})  . \label{3b13}
\eeq
The property that $\Gamma(q;P)$ is even in $P$ (and hence is real) has been
used.  The quantity $V(P;Q)$ is exactly $Q_{\nu}\Lambda_{\nu}(P;Q)$ due
to the Ward-Takahashi identity (\ref{3a5}).  The generated change in the
free action can therefore be written
\beqar
\delta \hat{S}_{2}[\pi] &=&
\frac{q_{\pi^{-}}}{(2\pi)^{2}}\int d^{4}(P,Q) \pi^{*}(P_{+})
\Lambda_{\nu}(P;Q) \delta A_{\nu}(Q) \pi(P_{-}) \nonumber \\
&=& \int d^4 Q J_{\nu}(Q) \delta A_{\nu}(Q)
, \label{3b14}
\eeqar
where we have constructed $\delta A_{\nu}(Q)=iQ_{\nu}\theta(Q)$.

Thus the gauge change in the total action $\hat{S}_{2}[\pi] -J \cdot A$
is zero and gauge invariance to first order is explicit.  The implication of
this analysis is that if the gauge transformation properties of the internal
$\bar{q}q$ structure factor of the bilocal pion field were ignored, then
(\ref{3b13}) for $V(P;Q)$ would become $\Delta^{-1}(P_{+}^2)-\Delta^{-1}
(P_{-}^2) \neq Q_{\nu}\Lambda_{\nu}(P;Q)$.  This becomes the point limit
on the mass-shell and gauge invariance for composite pions cannot be
maintained.  Although the free action for composite pions can be written
in the effective local form $\int d^4 P \pi^{*}(P) \Delta^{-1}(P^2)\pi(P)$,
it is not possible to gauge this action by minimal substitution and also
accommodate the distributed charge form factor for composite pions.  It is
necessary to recognize the substructure field content of $\Delta^{-1}(P^2)$
and develop the additional response to a gauge change that this introduces.

We note that the equivalence between the two expressions (\ref{3b5})
and (\ref{3b6}) for the gauge change of the composite pion field
implies a Ward-Takahashi identity for the electro-absorption vertex.
If we denote the $\bar{q}q$ vertex
of the pion in flavor channels $\alpha=(+,-)$ by
$\Gamma^{\alpha}(q;P)=i\gamma_{5}f_{\alpha}^{\dagger}\Gamma(q;P)$,
then the Ward-Takahashi identity satisfied by the
corresponding four-point vertex $\Gamma_{\nu}^{\alpha}(q;P,Q)$ is
obtainable from (\ref{3b5}) and (\ref{3b6}) in the form
\beq
Q_{\nu}\Gamma_{\nu}^{\alpha}(q;P,Q)= \hat{Q}\Gamma^{\alpha}
(q-\frac{Q}{2};P-Q) - \Gamma^{\alpha}(q+\frac{Q}{2};P-Q)\hat{Q}
- \left[\hat{Q}, \Gamma^{\alpha}(q;P-Q)\right] . \label{3b8a}
\eeq
This is one of the manifestations of gauge invariant photon coupling
to composite pions.

\section{Vector-Meson Dominance}

Although we have limited our considerations to electromagnetic coupling
to pions,
the technique can be extended to treat other meson components of the
fluctuation fields $\hat{{\cal B}}^{\phi}$ that appear
in the action (\ref{2b18}).  Of particular
importance for electromagnetic couplings are the neutral vector mesons
($\rho$ and $\omega$ in the present two flavor model) that implement
the successful vector dominance concept.  Consider the
neutral vector meson intermediate state contributions to the
photon-pion vertex. The usual dynamical signature for
vector dominance is the emergence of the current-field identity in
which the hadronic electromagnetic current is equivalent to a linear
combination of neutral vector meson fields.  That is, a bilinear
combination of the photon and vector mesons should appear in the action.  An
illustration of this is found in the NJL model where an exact current-field
identity holds.\cite{volkov82}
However, the lowest order occurrence of vector meson modes in the action
$\hat{S}[ \hat{{\cal B}}^{\phi },A_{\nu }]$ of (\ref{2b18})
is quadratic.  The formulation we have pursued does not immediately
yield a term that is bilinear in $A_{\nu}$ and $\hat{{\cal B}}^{\phi}$.
Nevertheless it would be incorrect to simply add a vector dominance
mechanism to the photon-pion vertex yielded by the present formalism.

The vector dominance mechanism is already included in the dynamical
content of the photon-quark
vertex $\Gamma_{\nu}(q;Q)$ defined by the inhomogeneous
integral equation (\ref{3a1}).  There are
intermediate $\bar{q}q$ ladder states of vector character that generate
a propagator pole whenever the photon momentum $Q$ is such that the
homogeneous version of the equation has a solution. An illustration
is provided in Fig. 3.
In the vicinity of a $\bar{q}q$ resonance we have
\begin{equation}
\Gamma _{\nu }\left(q;Q\right) \approx \frac{\Omega _{\nu
}(q;Q)}{Q^{2}+M_{V}^{2}},\label{2c1a}
\end{equation}
where $\Omega _{\nu }(q;Q)$ is the residue at the pole, and satisfies
the equation
\begin{equation}
\Omega _{\nu }\left(q;Q\right) =
-\frac{4}{3}g^{2}\int \frac{d^{4}k}{(2\pi )^{4}}D(q-k)\gamma _{\mu
}G_{0}\left( k+\frac{Q}{2}\right) \Omega _{\nu }\left( k;Q\right) G_{0}
\left(k-\frac{Q}{2}\right) \gamma _{\mu }, \label{2c2}
\end{equation}
for $Q^2=-M_{V}^2$.
This is the ladder Bethe-Salpeter equation
for a vector $\bar{q}q$ bound state.
The tree-level isoscalar and isovector mesons are degenerate and the flavor
structure is implemented simply by the quark charge operator
$\hat{Q}$ which multiplies both $\Gamma_{\nu}$ and $\Omega_{\nu}$.
This is the finite range counterpart of the observation previously
made in the context of the contact NJL model where explicit calculation
\cite{blin88} exhibits the $\rho$ meson peak.  The resulting
pion charge form factor\cite{pcff} has the correct qualitative
behavior with a vector dominance peak generated by the $Q$ dependence of
$\Gamma_{\nu}(q;Q)$.

\section{Summary}

The composite pion modes of the finite range four-fermion GCM model
have been used to study several issues that arise in relation to
a hadronic EM current and its basis at the quark level.  A Ward-Takahashi
identity is obtained in explicit form because the model allows a
propagator for the composite pion to be identified.  When the dynamics
is cast into the form of a nonlocal action for effective
local pion fields, the role of the hadronic $\bar{q}q$ form factors
in maintaining
the gauge invariance of the EM coupling is developed.
The effective inverse propagator for the localized fields contains
the substructure information and thereby has a gauge transformation
property generated by the extended nature of the
bilocal field content.
This needs to be included along with the
standard gauge transformation of effective local pion fields in order
for the gauge change in the free pion action to be exactly cancelled
by that of the EM current term via the Ward-Takahashi identity.

The modification to the pionic $\bar{q}q$ vertex due to a gauge change
can be viewed as the longitudinal component of a pion-quark
electro-absortion vertex.  The involvment of this modified vertex
in the EM coupling to Bethe-Salpeter model pions has been observed
before.\cite{gross}  Here, however, there is no need to supplement
the sum of one-body currents of the impulse approximation by an additional
term\cite{gross} based on this modified vertex to achieve gauge invariance.
The sum of one-body currents is automatically gauge invariant here because the
necessary dynamical relations among the elements of the triangle diagram
are generated from gauge invariance at the bare quark level.
Explicit knowledge of such a four-point vertex
would be needed to treat electromagnetic coupling to interactions between
bilocal $\bar{q}q$ fields rather than just the kinetic term.  In
particular, the contributions to the EM current from previously
derived\cite{praschifka87,mesondata} couplings
such as $\rho \pi \pi$ and $\omega \rho \pi$  could be
addressed with extensions of the present approach.
In a linear sigma model format of the GCM where an extended
$\bar{q}q$ scalar is included, the same mechanism will generate a
photon-sigma-quark vertex.  A gauge invariant EM coupling to the
chiral quark-meson model baryon\cite{frank91} of the GCM will not be possible
without explicit inclusion of such a mechanism.

The pion charge form factor at mean field level
within the GCM includes several physical phenomena that are not addressed
by the contact NJL model.
The dressed quark-photon vertex $\Gamma _{\nu}(q;Q)$ and the quark
self-energy depend on the loop momentum and a characteristic length
scale should be near the pion size.  The pion Bethe-Salpeter
amplitudes also depend on the loop momentum here and operate on the same
length scale.  Whether these elements can combine with the vector
dominance singularity to produce the pion charge radius remains to be
seen.  The GCM can be used to introduce a dynamical quark self-energy that
is confining through the absence of a pole in the propagator for real $p^2$.
This removes a threshold singularity that can influence the charge radius
just as strongly as the vector meson pole.

It has recently
been argued\cite{jaffe} that even with confined quarks, the charge form factor
at low momenta can be influenced by a threshold singularity
related to an effectively free behavior of quarks.  This would be driven
by the bound state wavefunction character of the pion $\bar{q}q$ amplitudes
with a scale set by the effective binding energy of the pion.  This
emphasizes the need for field theory models that accomodate pion size.

%_____________________________________________________________________

\acknowledgements

This work was supported in part by the National Science Foundation
under Grant Nos. HRD91-54080 and PHY91-13117.

\unletteredappendix{Normalization}

The normalization of the pion-photon vertex is such that
\mbox{$\Lambda_{\nu}(P;Q=0)=2P_{\nu}$} on the mass shell.  We outline a
verification of this.  From (\ref{piver}), we require
\beq
\Lambda_{\nu}(P;0)= \int d^{4}q \Gamma^{*}(q;P)
\Lambda_{\nu}(P,q;0)\Gamma(q;P),
\eeq
where \mbox{$\Lambda_{\nu}(P,q;0)$}, the photon vertex with bilocal pion
fields, is seen from (\ref{J2}) to involve the quark-photon vertex
\mbox{$\Gamma_{\nu}(q_{+};0)$}.  This in turn is determined by the integral
equation (\ref{3a1}), which in this limit is equivalent to the Ward identity
\mbox{$\Gamma_{\nu}(q;0)=-\frac{\partial}{\partial q_{\nu}} G_{0}^{-1}(q)$}.
This enables the soft photon limit of the quark loop to be expressed as
the momentum derivative of the inverse propagator for the bilocal pions.
After use of the symmetry property that \mbox{$\Lambda_{\nu}(P;0)$} is odd
in $P$, it is straightforward to obtain
\beq
\Lambda_{\nu}(P;0) = \int d^{4}(q',q) \Gamma^{*}(q';P)
\left[ \frac{\partial}{\partial P_{\nu} } \Delta^{-1}(q',q;P) \right]
\Gamma(q;P).
\eeq
The internal momentum integrations produce \mbox{$\Delta^{-1}(P^2)$}, the
inverse propagator for the localized pion fields, and we obtain
\beq
\Lambda_{\nu}(P;0) = \frac{\partial}{\partial P_{\nu}} \Delta^{-1}(P^{2}) -
\lambda(P^2) \frac{\partial}{\partial P_{\nu}} \int d^{4}q
\Gamma^{*}(q;P)\Gamma(q;P).
\eeq
Here the second term has been simplified through introduction of the
eigenvalue $\lambda$ associated with $\Gamma$ and $\Delta^{-1}$ according
to (\ref{2b21a}).  In the on-mass-shell limit
\mbox{$\lambda(-m_{\pi}^2)=0$}, and the first term produces $2P_{\nu}$
since \mbox{$\Delta^{-1}(P^2)=(P^2+m_{\pi}^2)Z(P^2)$}.
%_____________________________________________________________________

\figure{
The EM current or vertex for the composite pions of the GCM at tree level.  The
momenta labels
illustrate Eqs.(30), (31) and (32) of the text, which contain both $q$ and
$\bar{q}$
contributions.  The dynamically dressed quark propagators, the dressed
photon-quark vertex and
the pion $\bar{q}q$ amplitudes are defined with a consistent ladder structure
that maintains
gauge invariance. \label{fig1}}

\figure{
The four-point vertex for the electro-absorption of a composite pion on a
quark.  The
longitudinal amplitude $W_{\nu }(q;P,Q)$ for this process is involved in the EM
gauge transformation
of the $\bar{q}q$ pion field because of the spatially extended nature.  The
corresponding four-point
vertex has a Ward-Takahashi identity given by Eq.(58). \label{fig2}}

\figure{
(a) The ladder structure of the dressed photon-quark vertex given in Eq.(38).
(b) The vector
$\bar{q}q$ intermediate state propagator pole that is generated for
$Q^{2}\approx -M_{V}^{2}$.
\label{fig3}}

\end{document}